\documentclass[12pt]{article}

\usepackage{PRIMEarxiv}
\usepackage[utf8]{inputenc}
\usepackage[T1]{fontenc}
\usepackage{cmbright}
\usepackage{hyperref}
\usepackage{url}
\usepackage{booktabs}
\usepackage{amsfonts}
\usepackage{nicefrac}
\usepackage{microtype}
\usepackage{lipsum}
\usepackage{fancyhdr}
\usepackage{graphicx}
\usepackage{ragged2e}
\usepackage[table]{xcolor}
\usepackage{float}
\usepackage{caption}
\graphicspath{{media/}}

\title{ON THE EFFECT OF SIZE ON THE KINETICS OF REACTIONS IN SOLUTIONS}
\author{
  M.DEDOLA, G. CASSARÀ-AIROLDI\\
  Department of Chemistry, Life Sciences and Environmental Sustainability, University of Parma,\\
  Parco Area delle Scienze 17 A, Parma, Italy \\
  \And
  LUDOVICO CADEMARTIRI \\
  Department of Chemistry, Life Sciences and Environmental Sustainability, University of Parma,\\
  Parco Area delle Scienze 17 A, Parma, Italy \\
  Author to whom correspondence should be addressed:\\ ludovico.cademartiri@unipr.it
}

\begin{document}
\maketitle

\RaggedRight 

\begin{abstract}

Reactions in solution require “contact” between the reagents. We can predict the rate at which reagents come into “contact” (at least in dilute conditions), but if the initial collision does not lead to reaction, what happens then?\par
The collision rates in solution-phase reactions are generally described (explicitly or implicitly) with the Smoluchowski equation. Unfortunately, that model describes coagulations in gases, not reactions in solutions. The model is memory-less, i.e., collisions are treated as random processes, unaware of each other in space and time.\par
The reality is that unreactive collisions create memory: particles (even molecules) “remember” they just collided, i.e, the probability of collision depends on how far back in time their prior collision happened. As we show here, this purely geometric and statistical fact is valid as long as their size is larger than the Kuhn length of their Brownian motion in solution. Under these conditions, particles in solution form, even in the absence of attractive interactions, relatively long-lived “clusters” kept together by the statistical unlikeliness of separating. \\
We show here through Brownian dynamics simulations that, as a result of this memory, the collision rates and the lifetimes of these clusters, differently from what predicted by Smoluchowski’s model, are proportional to the ratio between the radius of the colliders and the Kuhn length of their path in solution, with a coefficient close to unity!
\end{abstract}

\section{Introduction}

A leitmotif in the history of science is that, as experimenters engage with increasingly complex systems,  parts of the theoretical toolbox (and associated assumptions) becomes increasingly inadequate in explaining or predicting observations: the evolution of chemistry has led us to study the reactivity of increasingly large species (e.g., macromolecules \cite{art1}, biomolecules \cite{art2}, colloids \cite{art3}, nanoparticles \cite{art4}, and surfaces \cite{art5}) and yet, very fundamental aspects of it are still unpredictable (e.g., kinetics of nucleation \cite{art6,art7} and protein-ligand binding \cite{art8,art9}).\par

As old models become increasingly inadequate, “corrections” become popular: a plethora of models (e.g. “crowding effects” \cite{art10},  “anomalous diffusion” \cite{art11}, “non-Arrhenius behavior” \cite{art12}) are now used to account for the discrepancies between the observed and predicted kinetics usually through adjustable parameters, such as “steric factors” \cite{art13,art14}.\par

In our previous work we encountered one such discrepancy: we demonstrated that nanocrystals of PbS can grow by coalescence well below their melting point, and were able to model quantitatively and predictively the kinetics of this process. \cite{art15} The values we obtained for the rate constants implied activation energies that, while consistent with prior work on oriented attachment \cite{art16,art17,art18,art19}, were not believably small (68 kJ/mol). We then posited that the collisional frequencies we assumed to estimate the activation energies (based on an interaction-corrected Smoluchowski model) could be severely underestimated. \cite{art15} \par

Reaction rates are usually described as the products of a collision rate and the probability with which each collision can lead to a reaction. \cite{art20,art21} Most of the attention among chemists is absorbed by the probability term (usually an exponential Boltzmann factor on an “activation energy”, which should contain information about the “reaction intermediates” and the overall "difficulty" of the process \cite{art22,art23}). But it is in the estimation of the collisional frequencies where old assumptions are hidden in plain sight.\par

The most widely used kinetic theory among chemists (“transition state theory”, TST) does not consider explicitly collision frequencies by assuming the reaction is not diffusion-limited and therefore that a dynamic equilibrium establishes between the reagents and a transition state which has a constant concentration. Kinetics is then entirely governed, in theory, by (i) the vibrational frequencies of the transition state, which determine its conversion into product, and (ii) the height $\Delta G$ of the barrier that separates the reagents from the transition state. \cite{art22}\par

In solution collisional theory (SCT), the collision frequencies are usually quantified using Smoluchowski's coagulation model. Discrepancies between the observed and predicted rate constants are corrected by considering interactions, and empirical parameters like the "steric factor". Despite its popularity, Smoluchowski's model carries assumptions that are nearly universally invalid in reactive systems: (i) it does not account for many-body collisions \cite{art24}; (ii) it does not account for the shape of the object \cite{art24,art25}  (This assumption is likely addressed by the steric factor, as proteins, for instance, have a non-spherical structure and reactions occur at specific sites. Therefore, the reactants must be correctly oriented to react\cite{art26,art27};  (iii) it does not account for reaction-limited processes. \cite{art28} \par

Smoluchowski formulated his model to describe the kinetics of coagulation in the gas phase, a process he reasonably assumed to occur irreversibly, and in dilute, diffusion-limited conditions (i.e., a negligible activation energy) \cite{art29}. Under such assumptions every collision is two-body and is conclusive: the particles are annihilated, a new particle is created, the probabilities of encounter in the system are reset. Of course, this is far from the vast majority of reactions which are instead reaction-controlled. 

\begin{figure}[h]
    \centering
    \includegraphics[width=0.85\linewidth]{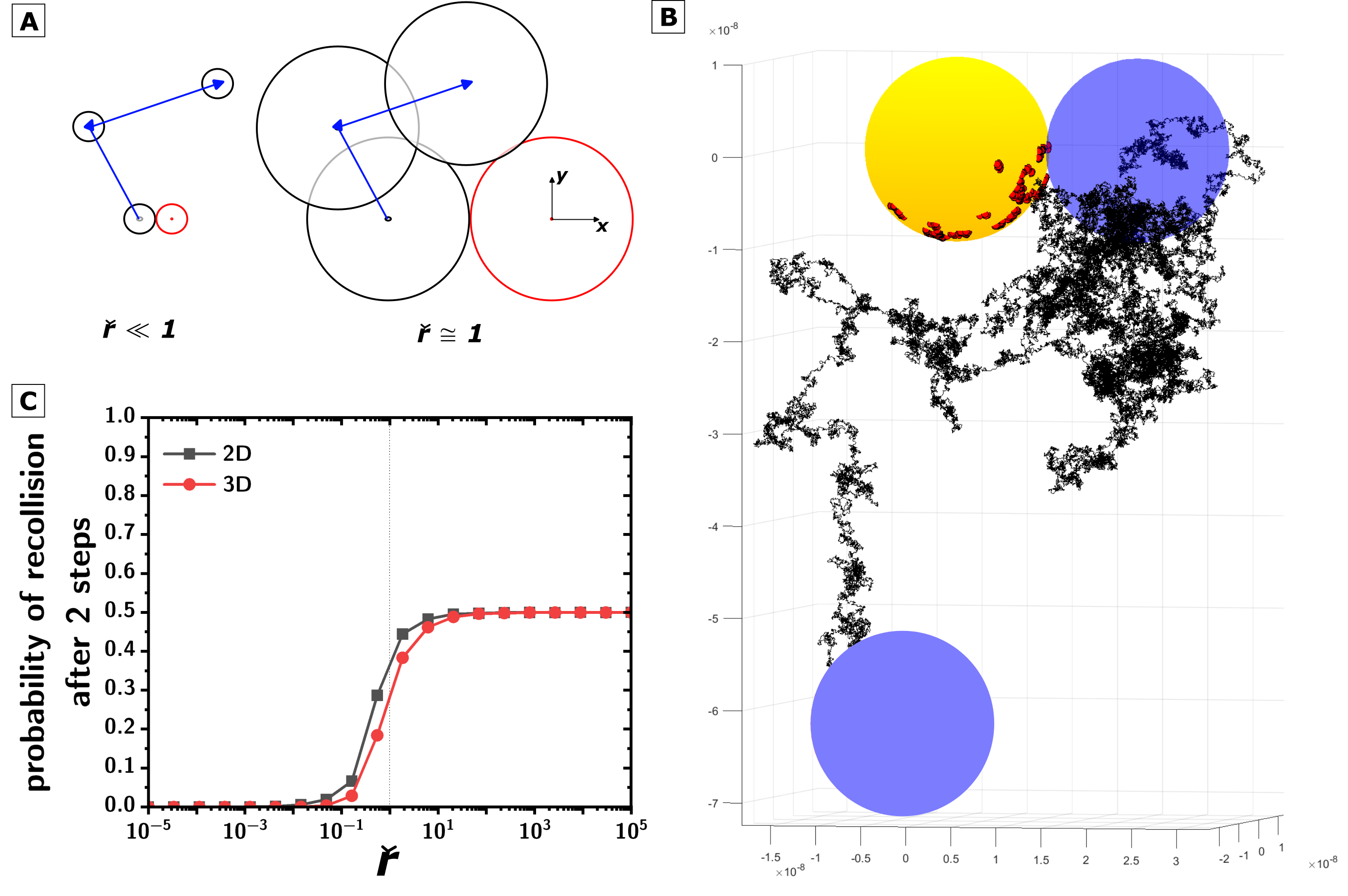}
    \caption{\textbf{Size-dependent correlation/memory in the collisional frequencies with constant Kuhn length.} \textbf{(A)}Diagram showing how the same two step path leads to no recollision or recollision depending on the size of the colliders. \textbf{(B)} Graph showing the Brownian motion (black line) of two particles (blue and gold), between a first collision and their separation by 6 radii. The displacement of the particles and the measurements of the axis are in meters. For clarity, the motion is plotted as that of the blue particle in the frame of reference of the golden particle. The red dots on the surface of the golden particle indicate the individual collisions that have occurred during this “encounter”. \textbf{(C)} Probability of recollision after two diffusion steps as a function of the relative size of the colliders (i.e., ratio between the particle radius and the mean free path) for 2D diffusion and 3D diffusion.}
    \label{Fig.1}
\end{figure}

In processes that are reaction-controlled we must focus on the most likely outcome: what happens if the first encounter is not successful. What does Brownian motions tells us?\par
The Brownian motion of objects in solution is a continuous trajectory with a persistence caused by inertia, is dissipated by friction with the solvent, and is caused (and decorrelated) by random collisions with it. These occur over times $\tau_c \sim \frac{r_s}{\sqrt{k_B T / m_s}}$, where $r_s$ is the radius of a solvent molecule, $m_s$ is its mass, $k_B$ is Boltzmann’s constant, and T is the temperature ($\tau_G \cong 10^{-13} \mathrm{~s}$ for $H_2O$ at room temperature). As a rule of thumb it is often stated that over timescales $\Delta t \gg \tau_c$ the dissipative effect of collisions is describable as an effective “viscous” force. \cite{art30} This is imprecise since it strongly depends on the effect of the average collision on the momentum vector of the Brownian object (the number of collisions required for dissipation will depend on the mass of the object and that of the solvent molecules). Under this assumption, the dynamic is then described as the effect of a net force consisting of two terms: a friction proportional to the particle velocity and a stochastic force (technically a “white noise”) that interprets the effect of thermal agitation. This turns out to be an Ornstein-Uhlenbeck process on the velocity of the particle which loses autocorrelation (both in magnitude and in orientation) exponentially in time with a decay constant (the “relaxation time”) equal (in the case of spheres of radius $r$ and uniform density $\rho$) to $\tau_r=\frac{2 \rho}{9 \eta} r^2$, where $\eta$ is the dynamic viscosity of the solvent.\par

An interesting observation is that this relaxation time plays a similar role in describing the trajectory of Brownian motion as the persistence length does in describing the conformation of worm-like chains. \cite{art31} The average correlation in the orientation of the tangent to two separate points on the worm-like chain is an exponentially decaying function of the contour distance between those two points: the persistence length is just the decay constant of that correlation decay function \cite{art32,art33}.\par

For mathematical simplicity, it is common to approximate persistent, continuous stochastic trajectories as discrete sequences of straight, uncorrelated steps. Kratky and Porod found that, for worm-like chains, the RMS end-to-end distance is equal to that of a non-persistent random walk with a step length equal to twice the persistence length \cite{art34}. This “Kuhn length” is therefore the shortest step length of a random walk that, on average, has the same conformation as the worm-like chain under consideration.\par

In the case of Brownian motion, we have the same exact problem but in time rather than in space: we know the decay constant of the decorrelation of motion in time, rather than the decay constant of the decorrelation of orientation of a chain in space. The mathematics that Kuhn employed applies identically here so one can take $2 \cdot \tau_r$  as the “Kuhn time” $\tau_K$ of the Brownian motion. Since $\tau_K>\tau_c$, we assumed for simplicity that the distance covered by the particle during the Kuhn time to be diffusive and we can call it “Kuhn step” $\lambda_K=\left(6 \mathrm{D}_{\mathrm{t}} \tau_K\right)^{1 / 2}$, where $D_t$ is the translational diffusivity of the particle. While this is an approximation (the step is not much longer than the relaxation time, so the motion over the timeframe $\tau_K$  is neither diffusive nor ballistic), the precise choice of the step length is not essential to the physical conclusions of this work, as long as the time associated with it is enough to cause near complete decorrelation (to assume memorylessness of the velocity vectors). Since this work aims to describe memory effects purely generated by the collisions, we had to choose time scales and length scales where the existing memory effects (i.e., persistence) would be negligible.

\begin{table}[h!]
\centering
\caption{Relaxation times $\tau_r$ and the Kuhn steps $\lambda_K$ for a range of molecules, clusters and colloids. Values are estimated at $T=298.15\,K$.}
\label{Tab1}
\begin{tabular}{|l|c|c|c}
\hline
\rowcolor{gray!20} 
\textbf{} & $\tau_{r}[s]$ & $\lambda_{K}[m]$ & $\check{r}=r/\lambda_K$ \\
\hline
EtOH & $9.64 \cdot 10^{-15}$ & $9.84 \cdot 10^{-12}$ & 25 \\
\hline
\rowcolor{gray!10} 
Au & $7.91 \cdot 10^{-14}$ & $3.90 \cdot 10^{-11}$ & 3 \\
\hline
CdSe (TOPO-capped, 5nm) & $5.27 \cdot 10^{-12}$ & $3.63 \cdot 10^{-11}$ & 97 \\
\hline
\rowcolor{gray!10} 
PbS (OLA-capped, 4nm) & $9.01 \cdot 10^{-12}$ & $4.43 \cdot 10^{-11}$ & 90 \\
\hline
Au (HDT-capped, 5nm) & $2.40 \cdot 10^{-11}$ & $7.15 \cdot 10^{-11}$ & 57 \\
\hline
\rowcolor{gray!10} 
SiO$_{\mathbf{2}}$ (2Å) & $4.90 \cdot 10^{-15}$ & $1.13 \cdot 10^{-11}$ & 9 \\
\hline
SiO$_2$ (20nm) & $4.90 \cdot 10^{-11}$ & $1.13 \cdot 10^{-10}$ & 88 \\
\hline
\end{tabular}
\end{table}

The Kuhn steps are usually under 1 Å for small nanoparticles or molecules (cf. Table \ref{Tab1}). They only exceed 1 Å for colloids (>20 nm) since, for constant density, $\lambda_k \propto r^{1 / 2}$. In summary, the motion of objects in solution can be described as an approximately non-persistent random walk whose steps are usually much smaller than the size of the object itself. \par
Our hypothesis is that, because of this simple fact, the collision rates should be much larger than those predicted by Smoluchowski’s model. For a constant $\lambda_K$ the probability two particles will recollide soon after the prior collision increases drastically with the size of the particles (cf. Figure \ref{Fig.1}A-C). This effect is not accounted for by the difference in cross-section (the Smoluchowski collision frequency for spheres is independent of their radius because it is proportional to the product of their diffusivity and their radius).\\
This enhanced probability of immediate recollision is a form of memory (i.e., the correlation of a system with an earlier state of self), which is reestablished every time two particles “recollide”: such “sticky” correlations can be quite significant on macroscopic averages (in this case collision frequencies). \par
It is important to note the memory mechanism we are talking about is neither inertia-based persistence in movement, nor the fluid-dynamic-based memory induced by the motion of the solvent around the diffusing particles. Both of these kinds of memory are well described already for the motion of single particles \cite{art35,art36,art37}. The memory mechanism we are discussing here is instead a collective emergent property of a system of particles. As far as we have been able to find, this mechanism has not been investigated quantitatively yet.

\section{Computational Design}
The design methodology was based on maximizing the statistics (which is necessary to prove unambiguously the existence of a memory effect and determine the decorrelation distance) as well as in using boundary conditions that minimize the influence of these effects (i.e., “worst case scenarios”). \par

\textbf{Code}. Three Brownian simulation codes were written independently by the three authors to ensure the robustness of the results.\par 

\textbf{Interactions}. The only interactions that were considered were hard-sphere step-wise potentials. Strong, long-range repulsive interactions would prevent collisions entirely at low volume fractions. Attractive interactions (even van der Waals) would only amplify the influence of the memory effects we are looking for. Long range interactions would also worsen the artifacts induced by periodic boundary conditions (e.g., motion correlation, action at a distance). Therefore the hard-sphere potential constitutes a “worst case scenario” for the hypothesis we are testing. Furthermore, it reduced the computational cost of the simulation and therefore allows for greater statistics.\par 

\textbf{Particle shape}. We chose to use spherical particles to have exact expressions of the translational diffusivity ($D_t=k_B T / 6 \pi \eta r$) and to simplify collision detection (Euclidean distance $< 2 \cdot r$). Any overlap would be considered a collision: there are no “active sites” on the surface of the particles. This assumption allows us to ignore rotational diffusion of the particles. All particles in the same simulation had the same radius. \par

\textbf{Step length}. The Brownian motion is described as a random flight where the individual steps are vector sums of normally distributed cartesian components with standard deviation equal to the Kuhn step. The viscosity $\eta$ value was chosen as $8.9 \cdot 10^{-4} Pa \cdot s$ . The temperature T was chosen as 298.15 K. 
The choice of step length is one of the most important for such simulations. The small step length helps avoiding artifacts and permit the choice of very small radii for the particles to assess the potential influence of these effects on molecular reactivity\par

\textbf{Type of collisions}. Considering the emphasis on simultaneously achieving high statistics and using extremely short step lengths, we decided to resolve collisions by returning the colliders to the positions they occupied at the beginning of the step. This approach allows to handle highly concentrated systems (volume fraction $\phi$ up to $10\%$), where each particle can collide multiple times in each time step, in a computationally effective manner. In such systems, the handling of collision chains in a fully elastic manner would require establishing a queue of movement within the step, or a further subdivision of the step in much smaller increments, which could introduce not necessarily predictable artifacts due to numerical instability \cite{art38}, as well as violate the underlying assumptions of describing solvent collisions as a net viscous force. \par 

\textbf{Number of collisions}. The simulations were run until a certain number of collisions were detected ($N_c$). The number was chosen on the basis of three considerations in this order of priority: (i) convergence of the average collision frequency; (ii) sufficient to extract distributions of intervals between collisions (“waiting times”) that allow to distinguish exponential (memory-less) from power-law (correlated) behavior; (iii) computational efficiency. In our case, the number of collisions that were collected, unless stated otherwise, was $\sim$1 million. \par

\textbf{Range of volume fractions}. One of the most important parameters in collisional dynamics is the volume fraction ($ \phi $). Low volume fractions ($\ll1\%$) allow to neglect multibody collisions and make the simulations closer to dilute conditions, but drastically increase computational times needed to achieve the desired statistics. Our strategy to address this conundrum was as follows. We used relatively high $ \phi $ (in 4 steps between $1\%$ and $10\%$) to (i) verify the convergence of the collision frequencies with $N_c$, (ii) verify reproducibility across codes and replicates, (iii) determine the number of particles N included in the sample volume that optimized the rate of collection of collisions. Once the convergence, reproducibility, and optimal N were determined at these relatively high $\phi$, all computational resources were concentrated on verifying that the observations at high $\phi$ would persist at very small $ \phi $ (we used $5\cdot 10^{-5} \%$ which corresponds approximately to a micromolar regime) while collecting $10^5$ rather than $10^6$ collisions. \par

\textbf{Range of particle radii}. The size of the particles was chosen to range between 1 Å and 10 nm in 10 logarithmically-spaced steps. Beside being unphysical, particles smaller than 1 Å would create very significant artifacts in the simulations (e.g., particles could go through each other without triggering a collision). Particles larger than 10 nm would diffuse very slowly and therefore cost too much from a computational standpoint without adding fundamentally new information. Large particles can also have $\lambda_K$ that are larger than the steps in our simulated random flights.\par 

\textbf{Boundary conditions}. The choice of boundary conditions, and the exact implementation in the code is very delicate. Periodic boundary conditions (PBC) using cubic unit cells are probably the most widely used approach for Brownian simulations and Molecular Dynamics \cite{art39,art40}. The most common implementation of them uses so-called “ghost particles” (cf. Figure \ref{Fig.S2}): as a particle intersect the boundary, ghost particles are created symmetrically at the opposite sides of the simulation volume to ensure the overall volume and surface area of the particles in the simulation volume stays constant. \\
While PBCs are significantly faster to simulate than large finite systems and avoid the edge effects they generate, they also come with limitations. (i) Ghost particles must be forced to move with the same displacement vector as the real particle that originated them to avoid notoriously persistent edge cases. This introduces correlations in the system, and, effectively, action at a distance. This problem is potentially ameliorated by introducing a sufficiently large number of particles in the simulation volume. (ii) The computational cost is somewhat elevated because up to 7 ghost particles (in the case of a cubic unit cell) can be simultaneously generated by a real particle and their positions calculated at each step. (iii) It constrains the size of the fluctuations of concentration in the system to be smaller than the unit cell.\\
To ensure that the artifacts introduced by PBCs would not be significant in our results, we used a closed system (“big box”, or BB) as a control (cf. Figure \ref{Fig.S2}). Specifically, if a PBC simulation involves a periodic unit cell of side length l, the corresponding BB control simulation would be a cubic closed system with side m·l where m is an odd number greater than 1. The rate constants in the BB control were calculated considering only the collisions occurring within the l x l x l “probe volume” sitting in the center of the BB system (cf. Figure \ref{Fig.S2}). Collisions with the outer walls were handled by spawning the particle in a random, unoccupied spot of the outer edge of the volume: this choice was aimed at reducing edge effects and eliminating “condensation” of particles at the walls. Theoretically, as m increases, the collision rates in the probe volume should converge towards the limit of the infinite system without incurring in the artifacts induced by PBC. In our simulations we used m ranging from 3 to 29 to verify the convergence of the collision frequency and calibrate with it the number of particles to use in the PBC simulation in order to match the observation in BB.\par

\textbf{Number of particles in the system}. As stated above, the BB conditions, in the limit of large m, should give the accurate value of collision frequencies. Figure \ref{Fig.S1} shows that the value of the second order rate constant k (i.e., the number of collisions observed in the sample volume per unit time divided by N2) converges with increasing m to approximately the same value to which k converges in PBC boundary conditions as N increases. This is not surprising in principle: as the boundary conditions are changed to mimic an infinite system more and more closely, the artifacts in the data diminish. We found it was sufficient to introduce $\sim$20 particles in PBC conditions to achieve a good approximation of the asymptote. To ensure this convergence would be preserved for all particle radii, we conducted full simulations (up to $N_c=1\cdot 10^{6}$) in PBC conditions for N ranging between 2 and 2001 in 10 logarithmically-spaced steps.\par 

\textbf{Ensuring convergence of the k values}. As shown in Figure \ref{Fig.S1}, the value of k calculated from $\sim 10^6$ collisions tends to converge as N increases (i.e., as the system becomes more statistically significant). On the other hand, the value of k should also ideally converge as the $N_c$ increases. To ensure we conduct our analysis on the most stable data, we analyzed the time series of k for all conditions we explored (N from 2 to 2001, $\phi $ from $5 \cdot 10^{-6}$ to 0.1, r from $1\cdot 10^{-10}$ m to $1\cdot 10^{-8}$ m), i.e., we calculated k at every time step considering the collisions occurred in the simulation until that step. The time series of k were then split in 10 time periods across the simulated time. For each of these time periods we conducted an Augmented Dickey-Fuller test (ADF) to determine whether that segment of time series had become stationary. If, upon reaching $10^{6}$ collisions, the value of k had not converged, the simulation was continued until convergence was obtained. Of the converged time series that remained, the most accurate k value was considered to be the ones at high N.

\section{Result and Discussion}
\begin{figure}[H]
    \centering
    \includegraphics[width=0.4\linewidth]{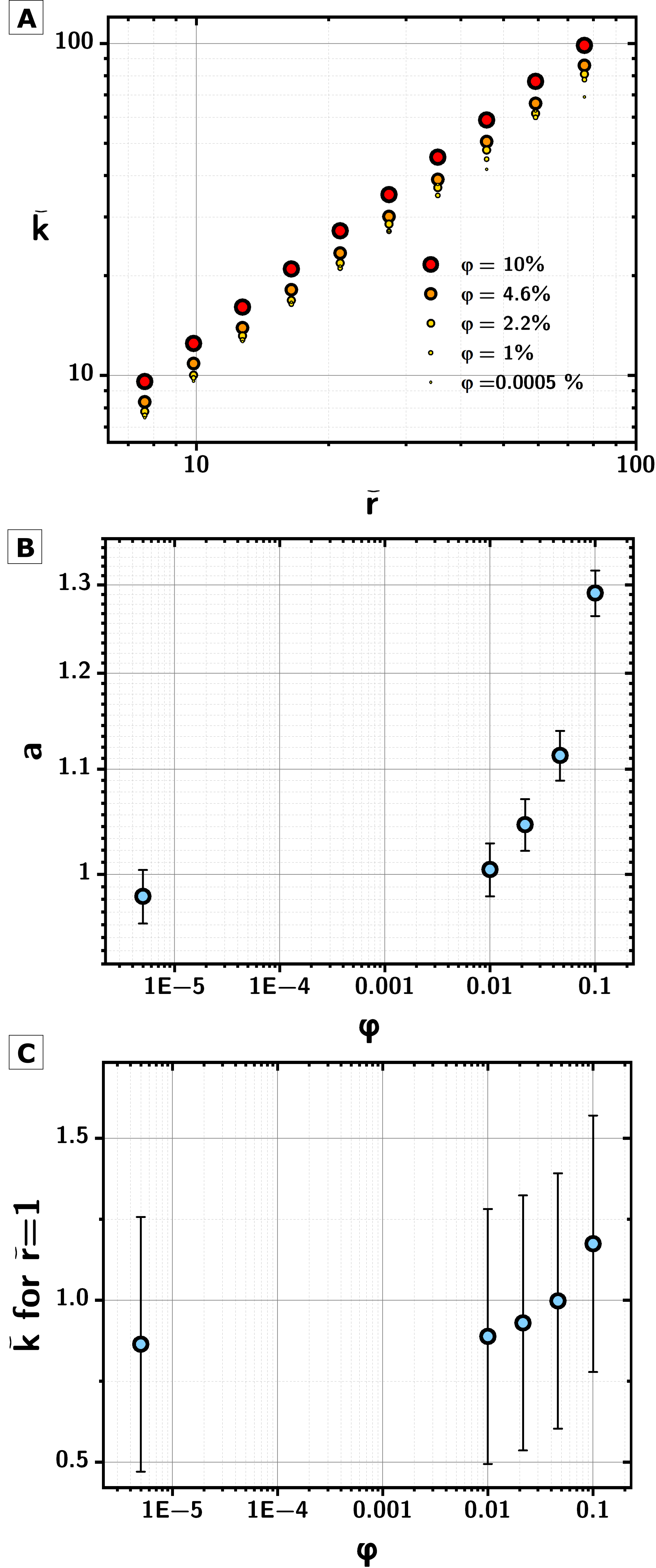}
    \caption{\textbf{Dependence of rate constants on the size of the colliders}. \textbf{(A)} Dependence of the ratio $\check{k}$ between simulated rate constant $k$ and the Smoluchowski's derived rate constant $k_S$ on the relative size $r$ of the colliders (i.e., $r / \lambda_k$ ). \textbf{(B)} Dependence of the slopes of the linear fits of data in (A) upon the volume fraction $\phi$. \textbf{(C)} Value of the linear fits in (A) for $\check{k}=1$.}
    \label{Fig.2}
\end{figure}

The ratio $\check{k}$ between the second order rate constants $k$ obtained from the simulations, and those predicted by Smoluchowski's model $k_{S}$ is proportional to $\check{r}$ across the entire range of $r$ we explored (cf. Figure \ref{Fig.2}A ). It is important to remind here the reader that the rate constant we describe here is, in the
conditions of our simulation (i.e., no reaction, or, in other words $e^{-E_a / k_B T} \sim 1$ where $\mathrm{E}_{\mathrm{a}}$ is the activation energy), identical to the so-called Arrhenius prefactor. \\
This linear trend $\breve{k}(\breve{r}, \varphi)=a(\varphi) \cdot \check{r}+b(\varphi)$ is conserved across all the values of $\varphi$ we characterized (the four values of "high $\varphi$ " between 1 and $10 \%$ for which $\mathrm{N}_c=10^6$, as well for $\varphi=5 \cdot 10^{-6}$ for which $\mathrm{N}_c=10^5$). The slopes $a(\varphi)$ of these linear trends converge to 1 as $\varphi \rightarrow 0$ (cf. Figure \ref{Fig.2}B). The best linear fits (with statistical weights of $1 / y$ ) gave values of $\check{k}$ that converge to $\sim 1$ as $\check{r} \rightarrow 1$ (cf. Figure \ref{Fig.2}C) with the exception of the highest $\varphi$.

This data allows us to draw some conclusions. \\
\begin{enumerate}
\item The dependence of $\check{k}$ over $\check{r}$ persists even when $\phi$ reaches very low values $\rightarrow$ the additional collisions are taking place after “first encounter” and are not due to a higher probability of the particles to find each other.
\item The dependence of $\check{k}$ over $\check{r}$ is preserved even when we use only two particles in the simulation volume (cf. Figure \ref{Fig.S3}) $\rightarrow$ the effect we observed is not caused by many-body collisions.\\
\end{enumerate}

\begin{figure}[H]
    \centering
    \includegraphics[width=0.4\linewidth]{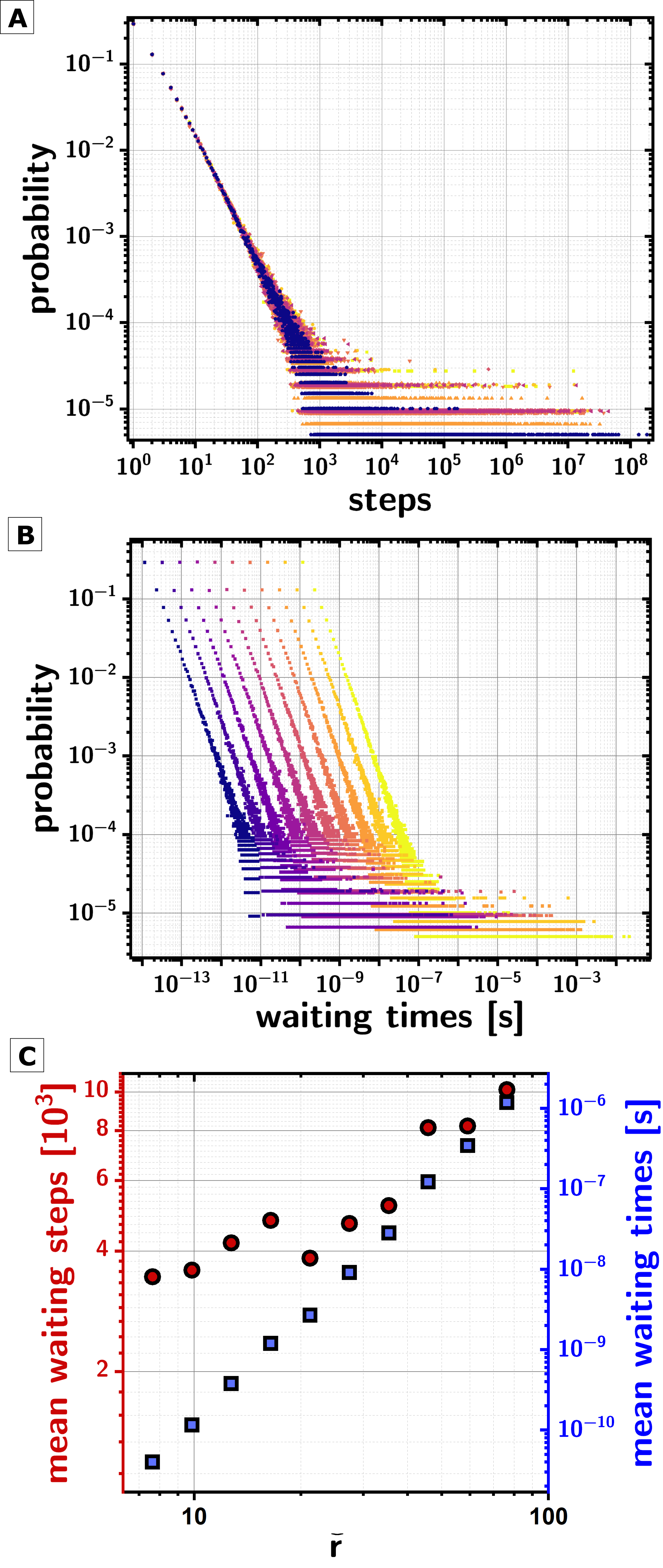}
    \caption{\textbf{Waiting times distributions}.\textbf{(A)} Distribution of “waiting steps” as a function of the relative size of the colliders in log-log plot. \textbf{(B)} Distribution of “waiting times” as a function of the relative size of the colliders in log-log plot. \textbf{(C)} mean waiting steps and waiting times as a function of $\check{r}$ .}
    \label{Fig.3}
\end{figure}

To better understand the origin of this effect we then looked at the distribution of “waiting” (i.e., the number of step or the period of times that separate successive collisions) and “residence” (i.e., the number of consecutive collisions or the time spent in consecutive collisions). 
In both cases we analyzed data from the lowest concentration ($\phi = 5 \cdot 10^{-6})$. In order to achieve sufficient statistics we filled the simulated volume with 400 labeled particles. The pairwise collisions were then separated for each pair of colliders. 
The distribution of “waiting” as a function of $\check{r}$ (cf. Figure \ref{Fig.3}) are rich with information. \\

\begin{enumerate}

\item The distribution of both “waiting steps” and “waiting times” appear to be a power law (cf. Figure \ref{Fig.3}A-B) with a very long tail that suggests the existence of a different distribution for very long times. 
\item The slope of the power law appears to be constant for all $\check{r}$.
\item While the distributions of the waiting steps is independent of $\check{r}$, the average waiting steps do change significantly with $\check{r}$, suggesting subtle changes in the tail of the distribution with $\check{r}$ (cf. Figure \ref{Fig.3}C)
\end{enumerate}

Smoluchowski’s model is expected to produce exponential “waiting times” as a consequence of the system “not remembering for how long it failed to produce an event” \cite{art41}. Exponential waiting times are characteristic of truly random, memory-less processes such as radioactive decay. The appearance of a power law is instead an indication of memory, i.e., that the collisions are not caused by entirely random processes or, in other words, that the collider still “carries information” about its past. \par

As the “waiting times” increase, the system should lose its memory of the prior collision and revert to randomness. Therefore it is expected that at long enough waiting times the distribution should revert to an exponential (test simulations we conducted using values of ř ~ 1 show very clearly the appearance of the exponential tail). In the conditions of our simulations though ($ \check{r}> 7$, $ \phi= 5 \cdot 10^{-6}$, $N = 400$, and $N_c=10^5$) we are unable to discern the exponential tail. This is likely because (i) the exponential trend occurs for waiting times corresponding to maximum distances between the particles that are larger than the size of the simulation volume, (ii) the statistics is not sufficient. In any case, it is remarkable to consider that our results show memory-rich waiting times for particles of molecular size and concentrations well into the dilute regime.\par 

The fact that average waiting steps and times increase with $\check{r}$ would seem to be at odds with the prior results that show an increase in the collision rate as a function of ř. This apparent contradiction can only be resolved if this effect is more than compensated by the time/steps that colliders spend in contact with each other (i.e., so-called “residence”), once the wait for a collision ends.\par 

\begin{figure}[H]
    \centering
    \includegraphics[width=0.4\linewidth]{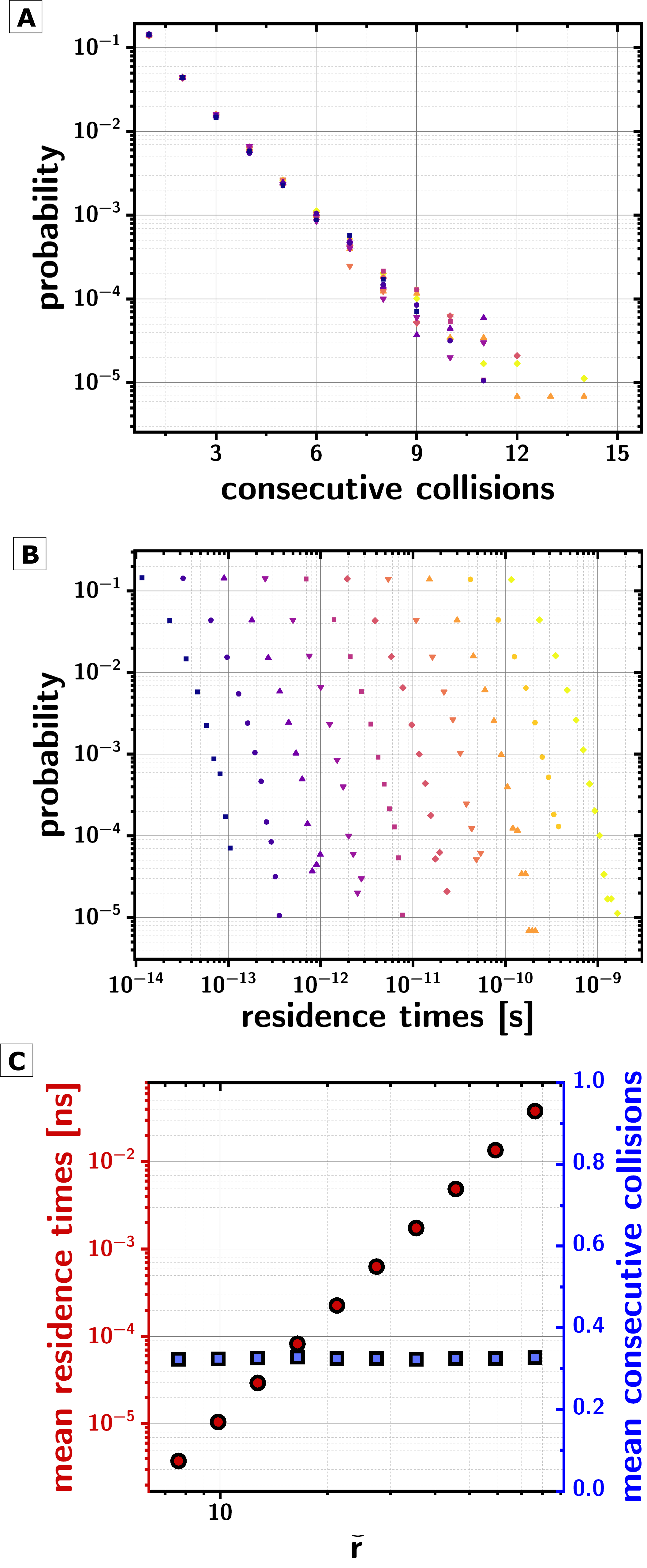}
    \caption{\textbf{Residence times}.\textbf{(A)} Distribution of consecutive collisions as a function of $\check{r}$ (blue to yellow) \textbf{(B)} Distribution of residence times as a function of $\check{r}$ (blue to yellow). \textbf{(C)} Mean values of the consecutive collisions and residence times as a function of $\check{r}$.}
    \label{Fig.4}
\end{figure}

To characterize “residence” we looked at the consecutive collisions between the same pairs of particles. The distribution of the “residence steps” and “residence times” (cf. Figure \ref{Fig.4}A-B) shows what look like exponential decays (they are not, in fact, exponential, having a longer tail but not long enough to be power laws). The non-exponentiality of the decay gave us some concerns that the average residence steps (and residence times) would not be “well-behaved”. Figure \ref{Fig.4}C shows that the average residence steps are independent of $\check{r}$. But when scaled in terms of time, due to the lower frequencies of steps performed by larger particles, the dependence of residence times vs radius is a power law. The exponent is very large (4.0016$\pm$0.0024) so that 200 nm particles showed residence times 4 orders of magnitude longer than 2 Å ones. If expressed in terms of r rather than ř, one obtains that the residence times are proportional to $r^2$ (e.g., 10 nm nanoparticles will spend 4 times more time in contact with each other than 5 nm particles).

\section{Conclusion}
In conclusion, we performed Brownian simulations on systems of hard spheres which showed that: 

\begin{enumerate}
    \item \textbf{Brownian dispersions of hard spheres show collision rate constants that are proportional to the square root of their radius rather than independent of them (as predicted by Smoluchowski’s model)}. The dependence is very strong (differences between observed and calculated through Smoluchowski’s model are up to 2 orders of magnitude) leading to a proposed correction to the collisional Smoluchowski prefactor $(A_s=16 \pi \cdot D_t \cdot r)$ between identical spheres of diffusivity $D_t$ to 

\begin{equation}
A=\frac{\eta A_{c s}}{\langle p\rangle_x} \cdot A_s
\end{equation}
    
Where $A_{cs}$ is the cross-sectional area of the collider, and $<p>_{x}$ is the RMS momentum of the particle on one coordinate. For identical spheres the entire collisional prefactor simplifies to 

\begin{equation}
A=4 \sqrt{\frac{k_B T \pi r}{\rho}}
\end{equation}

Our simulations indicate that this pseudoempirical expression should be valid for dilute solutions and for spherical particles/objects sizes ranging between 0.1 nm and 100 nm in radius, and show how the size can influence collisional kinetics. Our simulations employ no interactions. Attractive interactions (especially at short range) should only amplify such effects. \\
	\item \textbf{The observed collision rate appears to originate from a simple memory effect that was not considered by Smoluchowski due to his focus on coagulation processes rather than pure collisions}. This memory is such that probability of future collisions between particles depends on whether the particles have recently collided and failed to react. \\
 
\item \textbf{Objects significantly larger than the Kuhn length of their Brownian motion in solution can collide easily 10-100 times more than expected by commonly used collisional model, thereby introducing size as a significant variable in reactivity}. This factor could greatly influence the discovery and optimization of reaction paths from AI-enabled methods that rely in toto or in part on computationally-derived activation energies. 
\end{enumerate}

\newpage
\bibliographystyle{unsrt}  
\bibliography{references}  

\begin{thebibliography}{10}

\bibitem{art1}
B.~V. K.~J. Schmidt and C.~Barner-Kowollik.
\newblock Macromolecules made to order.
\newblock {\em Nature Chem}, 5(12), 2013.

\bibitem{art2}
Q.~Xue and E.~S. Yeung.
\newblock Differences in the chemical reactivity of individual molecules of an enzyme.
\newblock {\em Nature}, 373(6516):681--683, 1995.

\bibitem{art3}
Y.~Huang, C.~Wu, J.~Chen, and J.~Tang.
\newblock Colloidal self‐assembly: From passive to active systems.
\newblock {\em Angew Chem Int Ed}, 63(9):e202313885, 2024.

\bibitem{art4}
A.~Selmani, D.~Kovačević, and K.~Bohinc.
\newblock Nanoparticles: From synthesis to applications and beyond.
\newblock {\em Advances in Colloid and Interface Science}, 303:102640, 2022.

\bibitem{art5}
F.~Zaera, A.~J. Gellman, and G.~A. Somorjai.
\newblock Surface science studies of catalysis: Classification of reactions.
\newblock {\em Acc. Chem. Res.}, 19(1):24--31, 1986.

\bibitem{art6}
Y.~Diao, A.~S. Myerson, T.~A. Hatton, and B.~L. Trout.
\newblock Surface design for controlled crystallization: The role of surface chemistry and nanoscale pores in heterogeneous nucleation.
\newblock {\em Langmuir}, 27(9):5324--5334, 2011.

\bibitem{art7}
V.~Bhamidi, P.~J.~A. Kenis, and C.~F. Zukoski.
\newblock Probability of nucleation in a metastable zone: Induction supersaturation and implications.
\newblock {\em Crystal Growth \& Design}, 17(3):1132--1145, 2017.

\bibitem{art8}
Y.~Wang, J.~M. Martins, and K.~Lindorff-Larsen.
\newblock Biomolecular conformational changes and ligand binding: From kinetics to thermodynamics.
\newblock {\em Chem. Sci.}, 8(9):6466--6473, 2017.

\bibitem{art9}
B.~Schreier, C.~Stumpp, S.~Wiesner, and B.~Höcker.
\newblock Computational design of ligand binding is not a solved problem.
\newblock {\em Proc. Natl. Acad. Sci. U.S.A.}, 106(44):18491--18496, 2009.

\bibitem{art10}
J.~S. Kim and A.~Yethiraj.
\newblock Effect of macromolecular crowding on reaction rates: A computational and theoretical study.
\newblock {\em Biophysical Journal}, 96(4):1333--1340, 2009.

\bibitem{art11}
V.~Sposini, D.~Krapf, E.~Marinari, R.~Sunyer, F.~Ritort, F.~Taheri, C.~Selhuber-Unkel, R.~Benelli, M.~Weiss, R.~Metzler, and G.~Oshanin.
\newblock Towards a robust criterion of anomalous diffusion.
\newblock {\em Commun Phys}, 5(1):305, 2022.

\bibitem{art12}
W.~Wang and C.~J. Roberts.
\newblock Non-arrhenius protein aggregation.
\newblock {\em AAPS J}, 15(3):840--851, 2013.

\bibitem{art13}
S.~D. Traytak.
\newblock The steric factor in the time-dependent diffusion-controlled reactions.
\newblock {\em J. Phys. Chem.}, 98(31):7419--7421, 1994.

\bibitem{art14}
R.~D. Levine.
\newblock The steric factor in transition state theory and in collison theory.
\newblock {\em Chemical Physics Letters}, 175(4):331--337, 1990.

\bibitem{art15}
B.~Yuan and L.~Cademartiri.
\newblock Growth of colloidal nanocrystals by liquid‐like coalescence.
\newblock {\em Angew Chem Int Ed}, 60(12):6667--6672, 2021.

\bibitem{art16}
J.~Zhang, Y.~Wang, J.~Zheng, F.~Huang, D.~Chen, Y.~Lan, G.~Ren, Z.~Lin, and C.~Wang.
\newblock Oriented attachment kinetics for ligand capped nanocrystals: Coarsening of thiol-pbs nanoparticles.
\newblock {\em J. Phys. Chem. B}, 111(6):1449--1454, 2007.

\bibitem{art17}
C.~M. Evans, A.~M. Love, and E.~A. Weiss.
\newblock Surfactant-controlled polymerization of semiconductor clusters to quantum dots through competing step-growth and living chain-growth mechanisms.
\newblock {\em J. Am. Chem. Soc.}, 134(41):17298--17305, 2012.

\bibitem{art18}
J.~Zhang, Z.~Lin, Y.~Lan, G.~Ren, D.~Chen, F.~Huang, and M.~Hong.
\newblock A multistep oriented attachment kinetics: Coarsening of zns nanoparticle in concentrated naoh.
\newblock {\em J. Am. Chem. Soc.}, 128(39):12981--12987, 2006.

\bibitem{art19}
A.~L. Brazeau and N.~D. Jones.
\newblock Growth mechanisms in nanocrystalline lead sulfide by stopped-flow kinetic analysis.
\newblock {\em J. Phys. Chem. C}, 113(47):20246--20251, 2009.

\bibitem{art20}
S.~M. Blinder and C.~E. Nordman.
\newblock Collision theory of chemical reactions.
\newblock {\em J. Chem. Educ.}, 51(12):790, 1974.

\bibitem{art21}
K.~J. Laidler.
\newblock The meaning and use of the arrhenius equation.
\newblock {\em J. Chem. Educ.}, 61(6):494, 1984.

\bibitem{art22}
G.~A. Cook and A.~R. Ubbelohde.
\newblock The origin of the rate law.
\newblock {\em J. Chem. Educ.}, 41(9):456, 1964.

\bibitem{art23}
F.~Jensen.
\newblock Activation energies and the arrhenius equation.
\newblock {\em Quality \& Reliability Eng}, 1(1):13--17, 1985.

\bibitem{art24}
M.~B. Flegg.
\newblock Smoluchowski reaction kinetics for reactions of any order.
\newblock {\em arXiv}, 2015.

\bibitem{art25}
S.~I. Temkin and B.~I. Yakobson.
\newblock Diffusion-controlled reactions of chemically anisotropic molecules.
\newblock {\em J. Phys. Chem.}, 88(13):2679--2682, 1984.

\bibitem{art26}
V.~Gold.
\newblock {\em The IUPAC Compendium of Chemical Terminology: The Gold Book}.
\newblock International Union of Pure and Applied Chemistry (IUPAC), Research Triangle Park, NC, 4th edition, 2019.

\bibitem{art27}
M.~Schlosshauer and D.~Baker.
\newblock Realistic protein–protein association rates from a simple diffusional model neglecting long‐range interactions, free energy barriers, and landscape ruggedness.
\newblock {\em Protein Science}, 13(6):1660--1669, 2004.

\bibitem{art28}
M.~S. Johnson, J.~N. Mueller, C.~Daniels, H.~N. Najm, and J.~Zador.
\newblock Diffusion limited kinetics in reactive systems.
\newblock 2024.

\bibitem{art29}
M.~V. Smoluchowski.
\newblock Versuch einer mathematischen theorie der koagulationskinetik kolloider lösungen.
\newblock {\em Zeitschrift für Physikalische Chemie}, 92U(1):129--168, 1918.

\bibitem{art30}
J.~K.~G. Dhont.
\newblock {\em An Introduction to Dynamics of Colloids}.
\newblock Elsevier, Amsterdam, Netherlands, 1996.

\bibitem{art31}
M.~Rubinstein and R.~H. Colby.
\newblock {\em Polymer Physics}.
\newblock Oxford University Press, Oxford, New York, 2003.

\bibitem{art32}
H.-P. Hsu, W.~Paul, and K.~Binder.
\newblock Standard definitions of persistence length do not describe the local “intrinsic” stiffness of real polymer chains.
\newblock {\em Macromolecules}, 43(6):3094--3102, 2010.

\bibitem{art33}
M.~Doi and S.~F. Edwards.
\newblock {\em The Theory of Polymer Dynamics}.
\newblock International series of monographs on physics. Clarendon Press, Oxford, 1986.

\bibitem{art34}
O.~Kratky and G.~Porod.
\newblock Röntgenuntersuchung gelöster fadenmoleküle.
\newblock {\em Recl. Trav. Chim. Pays‐Bas}, 68(12):1106--1122, 1949.

\bibitem{art35}
T.~Franosch, M.~Grimm, M.~Belushkin, F.~M. Mor, G.~Foffi, L.~Forró, and S.~Jeney.
\newblock Resonances arising from hydrodynamic memory in brownian motion.
\newblock {\em Nature}, 478(7367):85--88, 2011.

\bibitem{art36}
F.~Donado, R.~E. Moctezuma, L.~López-Flores, M.~Medina-Noyola, and J.~L. Arauz-Lara.
\newblock Brownian motion in non-equilibrium systems and the ornstein-uhlenbeck stochastic process.
\newblock {\em Sci Rep}, 7(1):12614, 2017.

\bibitem{art37}
V.~Lisy and J.~Tothova.
\newblock On the (hydrodynamic) memory in the theory of brownian motion.
\newblock {\em arXiv}, 2004.

\bibitem{art38}
R.~C. Hilborn.
\newblock {\em Chaos and Nonlinear Dynamics: An Introduction for Scientists and Engineers}.
\newblock Oxford University Press, Oxford, 2nd edition, 2000.

\bibitem{art39}
Weidong Wu et~al.
\newblock Applying periodic boundary conditions in finite element analysis.
\newblock In {\em SIMULIA community conference}, Providence, 2014.

\bibitem{art40}
M.~Długosz, P.~Zieliński, and J.~Trylska.
\newblock Brownian dynamics simulations on cpu and gpu with bd\_box.
\newblock {\em J Comput Chem}, 32(12):2734--2744, 2011.

\bibitem{art41}
Maciej Dobrzynski et~al.
\newblock When do diffusion-limited trajectories become memoryless?
\newblock In {\em Proceedings of the 5th Workshop on Computation of Biochemical Pathways and Genetic Networks}, 2008.

\end{thebibliography}

\pagestyle{fancy}
\fancyhf{}
\fancyhead[LO]{SUPPORTING INFORMATION: ON THE EFFECT OF SIZE ON THE KINETICS OF REACTIONS IN SOLUTIONS}
\fancyfoot[C]{\thepage}

\begin{titlepage}
\centering
{\LARGE\bfseries SUPPORTING INFORMATION: ON THE EFFECT OF SIZE ON THE KINETICS OF REACTIONS IN SOLUTIONS \par}

\vspace{0.5cm}

{\large
\textbf{M.DEDOLA}, \textbf{G. CASSARÀ-AIROLDI} \\
Department of Chemistry, Life Sciences and Environmental Sustainability, University of Parma,\\
Parco Area delle Scienze 17 A, Parma, Italy \par
}

\vspace{0.3cm}

{\large
\textbf{LUDOVICO CADEMARTIRI} \\
Department of Chemistry, Life Sciences and Environmental Sustainability, University of Parma,\\
Parco Area delle Scienze 17 A, Parma, Italy \\
Author to whom correspondence should be addressed: \\
ludovico.cademartiri@unipr.it \par
}

\end{titlepage}

\RaggedRight

\setcounter{section}{0}
\section{Supporting Information}

\renewcommand{\thefigure}{S\arabic{figure}}
\setcounter{figure}{0}

\begin{figure}[H]
    \centering
    \includegraphics[width=0.6\linewidth]{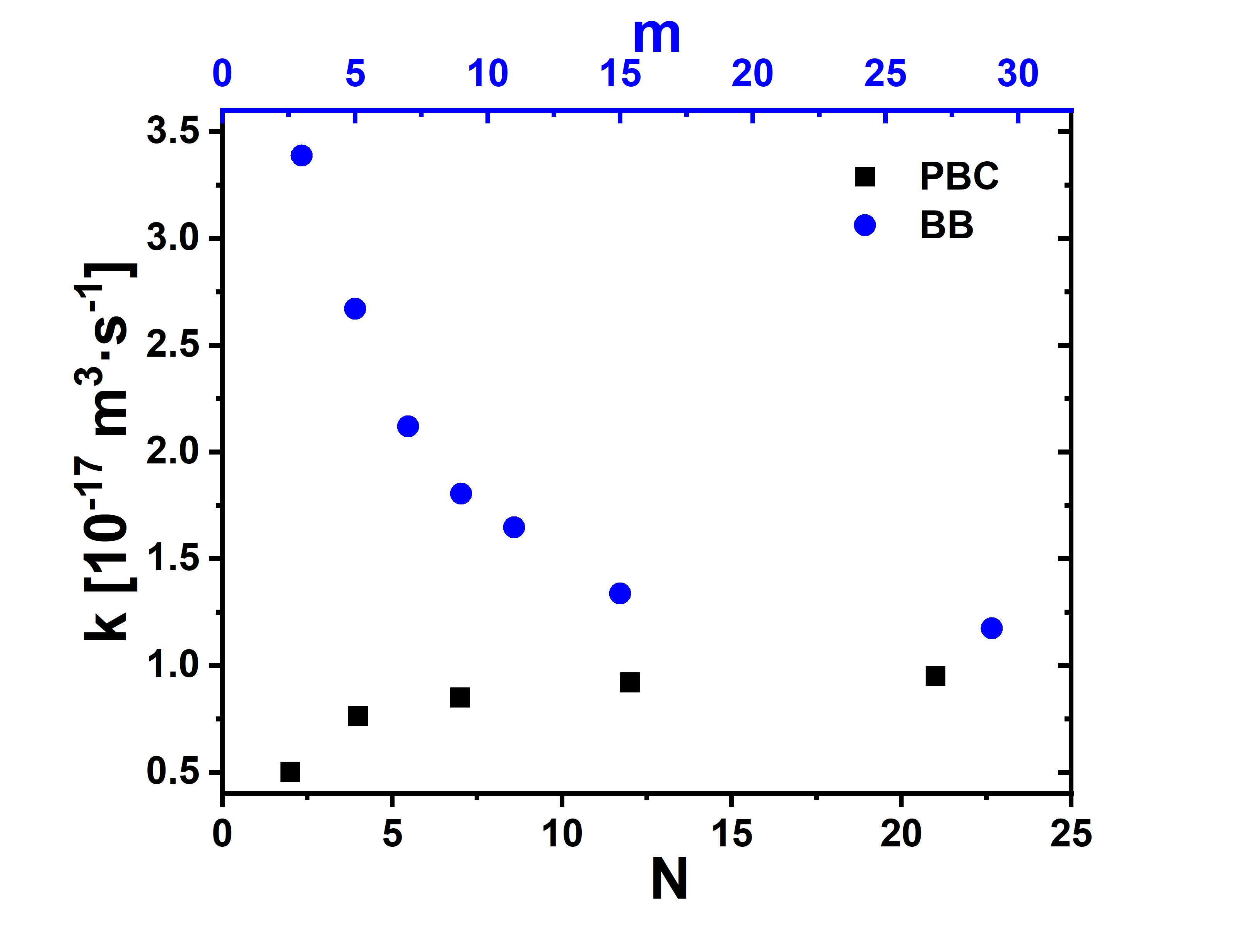}
    \caption{Collisional rate constants measured after $10^{6}$ collisions for hard spheres with radius $r=10^{-9}$ m, as a function of the number of particles examined within volumes defined by Periodic Boundary Conditions or Big Box conditions (cf. Figure \ref{Fig.S2}).}
    \label{Fig.S1}
\end{figure}

\begin{figure}[H]
    \centering
    \includegraphics[width=0.6\linewidth]{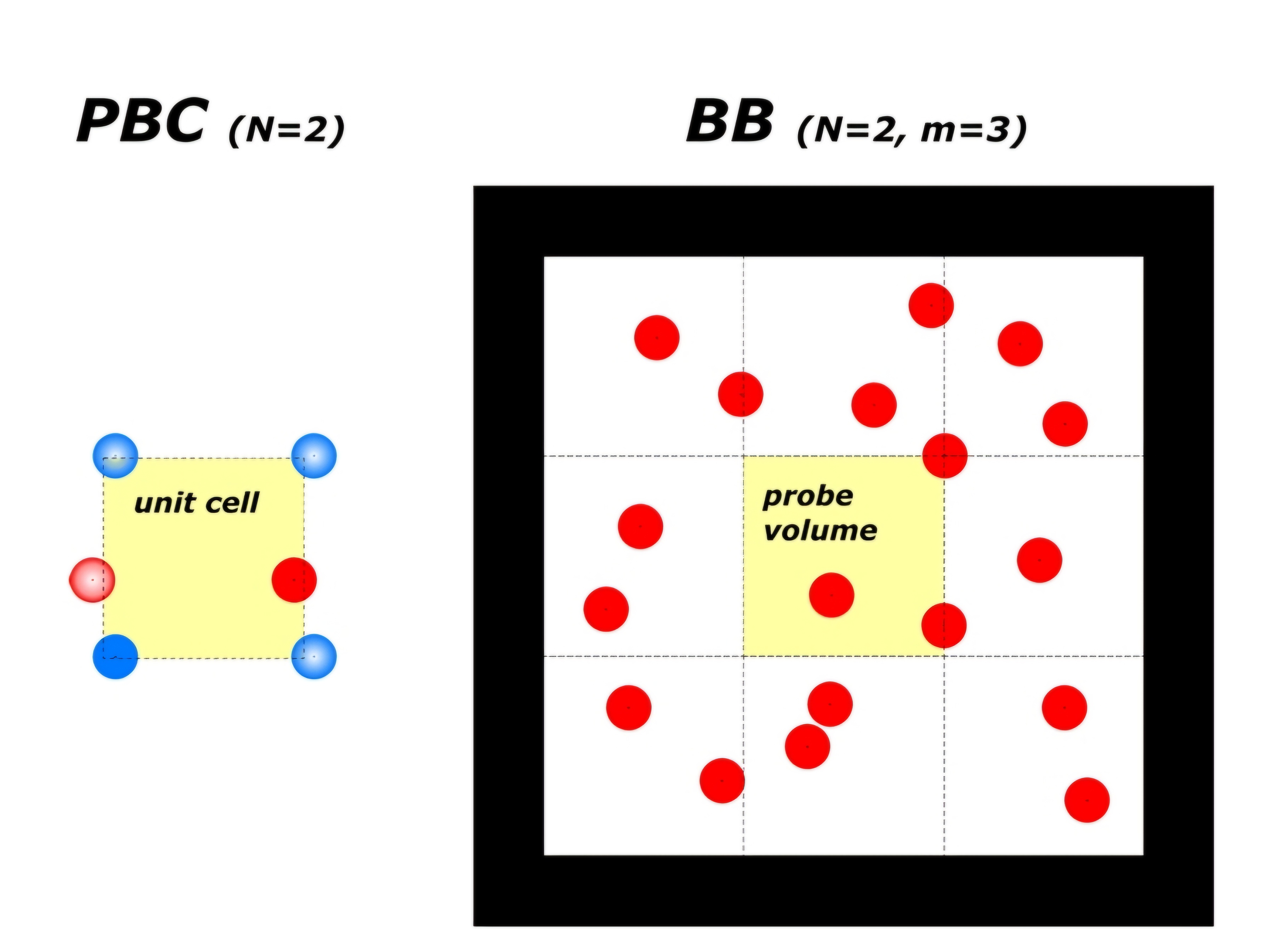}
    \caption{Sketch of the two boundary conditions used to evaluate the collisional rate constants.}
    \label{Fig.S2}
\end{figure}

\begin{figure}[H]
    \centering
    \includegraphics[width=0.50\linewidth]{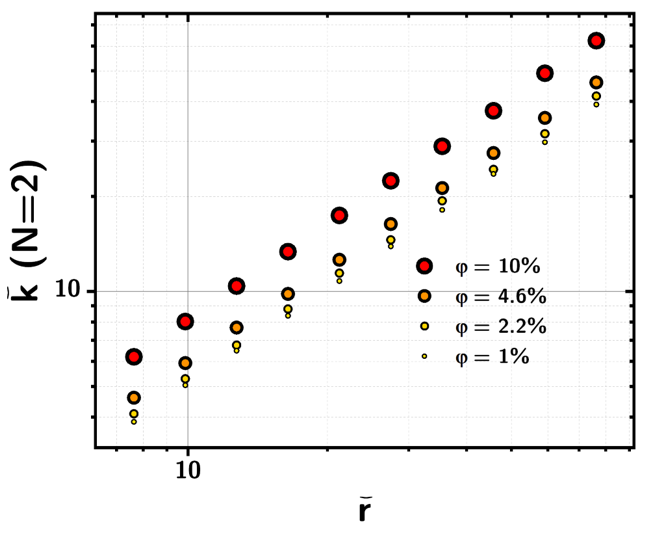}
    \caption{Dependence of $\check{k}$ over $\check{r}$ for a variety of $\phi$ when only two particles are present in the simulation volume (PBC, so all collisions are two-body collisions).}
    \label{Fig.S3}
\end{figure}

\end{document}